\documentclass[aps,prl,twocolumn,superscriptaddress,showpacs]{revtex4}
\usepackage{makeidx}
\usepackage{graphicx}
\usepackage{amsmath}
\usepackage{amsfonts}
\usepackage{amssymb}
\usepackage{bm}
\usepackage{color}
\usepackage[varg]{txfonts}

\def\Im{\mathop{\rm Im}}

\def\be{\begin{equation}}
\def\ee{\end{equation}}
\def\Landau{\alpha}
\def\fl{{\mathrm{fl}}}

\begin{document}

\title{Giant Nernst Effect due to Fluctuating Cooper Pairs in Superconductors}
\author{M. N. Serbyn}
\affiliation{Landau Institute for Theoretical Physics, Chernogolovka, Moscow
Region, 142432, Russia}
\author{M. A. Skvortsov}
\email{skvor@itp.ac.ru}
\affiliation{Landau Institute for Theoretical Physics, Chernogolovka, Moscow
Region, 142432, Russia}
\author{A. A. Varlamov}
\affiliation{COHERENTIA-INFM, CNR, Viale del Politecnico 1, I-00133 Rome, Italy}
\author{Victor~Galitski}
\affiliation{Joint Quantum Institute and CNAM, Department of Physics, University of
Maryland, College Park, MD 20742-4111, USA}
\date{June 27, 2008}

\begin{abstract}
A theory of the fluctuation-induced Nernst effect is developed for arbitrary
magnetic fields and temperatures beyond the upper critical field line in a
two-dimensional superconductor. First, we derive a simple phenomenological
formula for the Nernst coefficient, which naturally explains the giant
Nernst signal due to fluctuating Cooper pairs. The latter is shown to be
large even far from the transition and may exceed by orders of magnitude the
Fermi liquid terms. We also present a complete microscopic calculation
(which includes quantum fluctuations) of the Nernst coefficient and give its
asymptotic dependencies in various regions on the phase diagram. It is
argued that the magnitude and the behavior of the Nernst signal observed
experimentally in   disordered superconducting films can be
well-understood on the basis of the superconducting fluctuation theory.
\end{abstract}


\maketitle

A series of recent experimental studies has revealed an anomalously strong
thermomagnetic signal in the normal state of the high-temperature
superconductors~\cite{Ong1,Ong2,Ong3,hTc1,hTc2,hTc3,Ong4,Ong5,Rick} and
disordered superconducting films~\cite{Aubin1,Aubin2}. In the pioneering
experiment \cite{Ong1}, Xu \emph{et al}.\
observed a sizeable Nernst effect in the La$_{2-x}$Sr$_{x}$CuO$_{4}$
compounds up to $130$~K, well above the transition
temperature, $T_{c}$. This and further similar experiments on the cuprates
have sparked theoretical interest in the thermomagnetic phenomena.
Theoretical approaches to the anomalously large Nernst-Ettingshausen effect
currently include models based on the proximity to a quantum critical point~%
\cite{Sachdev}, vortex motion in the pseudogap phase~\cite%
{Ong2,Ashvin,Ashvin_Huse}, as well as a superconducting fluctuation scenario~%
\cite{Uss1,Uss2,Uss3}. While the two former theories are specific to the
cuprate superconductors, the latter scenario should apply to other more
conventional superconducting systems as well. Very recently, a large Nernst
coefficient was observed in the normal state of disordered superconducting
films~\cite{Aubin1,Aubin2}. These superconducting films are likely to be
well-described by the usual BCS model and, hence, the new experimental
measurements provide indication that the superconducting fluctuations
are likely to be the key to
understanding the underlying physics of the giant thermomagnetic response.

Various groups have previously calculated the fluctuation-induced Nernst
coefficient in the vicinity of the classical
transition~\cite{Dorsey,ReizerSergeev1994,Uss1,Uss2,Uss3}.
However, these analyses were limited to the case of
very weak magnetic fields and temperatures close to the zero-field
transition, when Landau quantization of the fluctuating Cooper pair motion
can be neglected.
In experiment, however, other parts of the phase diagram (in particular
strong fields) are obviously important and how the quantized motion of
fluctuating pairs would figure into the thermomagnetic response has remained
unclear. In this Letter we clarify this physics, explaining the origin of
the giant fluctuation Nernst-Ettingshausen effect, and develop a complete
microscopic theory of Gaussian superconducting fluctuations at
arbitrary magnetic fields and temperatures.

We start with a qualitative discussion of the Nernst-Ettings\-ha\-u\-sen effect.
Consider a conductor in the presence of a magnetic field, $H_z$, and electric
field, $E_y$, directed along the $z$- and $y$-axes respectively.
The charged carriers subject to these crossed fields acquire a drift velocity
$\overline{v}_x=cE_y/H_z$ in the $x$-direction. The latter would
result in the appearance of a transverse current
$j_x = ne\overline{v}_x$.
When the circuit is broken, no current flows,
and the drift of carriers is prevented by the spacial variation
of the electric potential:
$\nabla_x\varphi=-E_x=(nec/\sigma)(E_y/H_z)$,
where $\sigma$ is the conductivity.
Due to electroneutrality, this generates the spacial
gradient of the chemical potential:
$\nabla_x\mu(n,T) + e\nabla_x \varphi=0$,
which corresponds to the appearance of the
transverse temperature gradient $\nabla_x T = \nabla_x\mu(d\mu/dT)^{-1}$
along the $x$-direction.
Hence, the Nernst coefficient can be expressed in terms of the full
temperature derivative of the chemical potential:
\be
  \nu_N
  =
  \frac{E_y}{(-\nabla_x T)H_z}
  =
  \frac{\sigma}{ne^{2}c}
  \frac{d\mu}{dT}
  .
\label{nernstdef}
\ee

E.g., in a degenerate electron gas, the chemical potential
$\mu(T) = \mu_0 - (\pi^2T^2/6)(d\ln\nu/d\mu)$,
where $\nu(\mu)$ is the density of states,
and one easily reproduces the value of the Nernst coefficient in a normal metal:
$\nu_N=(\pi^2 T/3mc) (d\tau/d\mu$)
\cite{Sondheimer1948,footnote},
where $\tau$ is the elastic scattering time (here and below $\hbar=k_{B}=1$).
Thus the Nernst effect in metals is small
due to the large value of the Fermi energy.

The simple form of Eq.~(\ref{nernstdef}),
which to the best of our knowledge has not appeared
previously in the literature, suggests that in order to get a large Nernst
signal, \emph{a strong temperature dependence of the chemical potential of
carriers is required}. This can be achieved in the vicinity of the
superconducting transition where
a new type of carriers
(fluctuating Cooper pairs) appear besides the normal electrons. These
excitations are unstable, have the characteristic lifetime of order
$\tau_{\mathrm{GL}}=\pi /8(T-T_c)$,
and form an interacting Bose gas
with a variable number of particles. In two dimensions, their concentration
is $n_{\mathrm{c.p.}}^{(2)}(T) = (m T_c/\pi) \ln[T_c/(T-T_c)]$
\cite{LV}. In the vicinity of $T_{c}$, the
chemical potential of fluctuating Cooper pairs can be found by
identifying its value in the Bose distribution to give
$n_{\mathrm{c.p.}}^{(2)}(T)$ above.
This leads to $\mu_{\mathrm{c.p.}}^{(2)}(T) = T_{c}-T$.
Since $d\mu_{\mathrm{c.p.}}^{(2)}/dT=-1$,
the fluctuation contribution to the Nernst signal (\ref{nernstmagn})
exceeds parametrically the Fermi liquid term. In this
sense it is similar to the fluctuation diamagnetism (which also exceeds the
Landau/Pauli terms and is effectively a correction to the perfect
diamagnetism of a superconductor). Based on Eq.~(\ref{nernstdef}) and using
the known expression for paraconductivity in a magnetic field,
$\sigma_{\fl}=(e^2/2\epsilon)F(\epsilon/2\tilde h)$ \cite{LV}, one
finds (with logarithmic accuracy) the value of the Nernst coefficient due to
fluctuating Cooper pairs:
\begin{gather}
  \nu_N^{(2)}(T,H)
  \sim
  \frac{1}{mc}
  \frac{F(x)}{T-T_{c}}
  \sim
\begin{cases}
[mc(T-T_{c})]^{-1}, & x\gg 1, \\
(meDH)^{-1}, & x\ll 1,
\end{cases}
\label{nernstmagn}
\\
  F(x)
  =
  x^2 \left[ \psi \left( 1/2+x\right)
  - \psi \left( x\right) -1/(2x) \right] ,
\label{F(x)}
\end{gather}%
where $x=\epsilon/2\tilde{h}$,
$\epsilon=\ln(T/T_{c})$ and $\tilde{h}=H/\widetilde{H}_{c2}(0)$
are the reduced temperature and
magnetic field, $\widetilde{H}_{c2}(0)=4cT_{c}/\pi eD$ is the
linearly extrapolated value of the upper critical field, and
$D$ is the diffusion coefficient.
The estimate (\ref{nernstmagn})
corresponds to the results of Ginzburg-Landau (GL)~\cite{Uss1,Dorsey} and
diagrammatic \cite{Uss2} treatment in the vicinity of the classical transition.

To develop a quantitative theory of observable thermoelectric transport,
we have to recall a deep relation between the fluctuation Nernst effect and
magnetization as emphasized in Refs.~\cite{diaNernst,CHR97,Uss1}:
In the presence of a magnetic field, the measurable transport heat current
$\mathbf{j}^Q_\text{tr}$ differs from the microscopic heat current
$\mathbf{j}^Q$ by the circular magnetization current
$\mathbf{j}^Q_M=c\mathbf{M}\times\mathbf{E}$,
where $\mathbf{M}$ is the induced magnetization.
As a result, the thermoelectric tensor $\beta^{\alpha\beta}$
relating $j^{Q\alpha}_\text{tr}=T\beta^{\alpha\beta}E^\beta$
with the applied electric field $\mathbf{E}$
can be found as a sum of the kinetic, $\tilde\beta^{\alpha\beta}$,
and thermodynamic, $\beta_M^{\alpha\beta}$, contributions:
\be
  \beta^{\alpha\beta}
  =
  \tilde\beta^{\alpha\beta}
  +
  \beta^{\alpha\beta}_M,
\qquad
  \beta^{\alpha\beta}_M
  =
  \epsilon^{\alpha\beta\gamma} c M^\gamma / T .
\label{betagen}
\ee
The term $\tilde\beta^{\alpha\beta}$ can be expressed
via the thermal and quantum mechanical averaging
of the electric-heat currents correlator
$Q^{\alpha\beta}(\omega_\nu)
= \bigl\langle j^{e\alpha}(-\omega_\nu) j^{Q\beta}(\omega_\nu) \bigr\rangle$
by analytic continuation to real frequencies:
$
  \tilde{\beta}^{\alpha\beta}
  =
  T^{-1}
  \lim_{\omega \rightarrow 0}
  \Im Q^{\alpha\beta} (-i\omega+0) /\omega
$,
while the term $\beta_M^{\alpha\beta}$
accounts for the magnetization heat current $\mathbf{j}^Q_M$.

The general expression \cite{Kurkijarvi,LV} for the
fluctuation magnetization $M(T,H)$
has been calculated previously in the GL
region~\cite{Klemm,LV} and at low temperatures, close to
$H_{c2}(0)$~\cite{VGL}.

\begin{figure}
\includegraphics[width=.9\columnwidth]{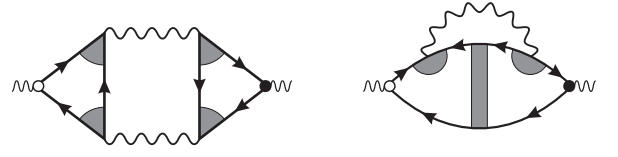}
\caption{The Aslamazov-Larkin (AL) and density-of-states
(DOS) diagrams for the thermoelectric response $\tilde\beta^{xy}$.
The DOS diagram has a symmetric counterpart.
The white and black circles
correspond to the different heat and electric vertices,
the shadowed blocks represent cooperons,
and the wavy lines denote the fluctuation propagator (see text).
All objects on these graphs are generally matrices in the Landau basis.}
\label{figa1}
\end{figure}

Now we proceed with the microscopic calculation of the Nernst coefficient
$\nu_N(T,H) = R_\Box\beta^{xy}/H$ at arbitrary $T$ and $H$.
We concentrate on the calculation of the linear response operator
$Q^{xy}(\omega_\nu)$ and its
analytic continuation to real frequencies.
In order to get the most general expression
we follow Ref.~\cite{VGL} and perform
calculations in the Landau basis, without expanding the Green's functions,
propagators, current and heat vertices in the magnetic field. Within this
approach, the fluctuation part of the correlator $Q^{xy}(\omega_\nu)$
is generally represented by ten diagrams\ \cite{VGL,LV}.
However, in the case of the Nernst effect, the
Maki-Thompson contribution to Nernst coefficient can be shown to be exactly
zero. What concerns the positive Aslamazov-Larkin (AL) term, it dominates in
the Ginzburg-Landau (GL) region in the immediate vicinity of the classical
transition and competes with the negative density-of-states (DOS)
contribution everywhere else. It is interesting that the hierarchy of the DOS
diagrams in the problem under consideration in unusual: Only the graphs
containing three cooperons (see Fig.~\ref{figa1}) turn out to be important.
The AL and DOS contributions, and the fluctuation magnetization can be written as
\begin{gather}
Q_{\text{AL}}^{xy}(\omega_\nu)
=
-4 \nu_H T \sum_{\Omega _{k}}\sum_{n,m}%
\hat{q}_{mn}^{x}\mathcal{B}_{nm}^{(e)}\hat{q}_{nm}^{y}\mathcal{B}%
_{nm}^{(Q)}L_{m}(\Omega _{k+\nu })L_{n}(\Omega _{k}),
\label{AL}
\\
  2 Q_\text{DOS}^{xy}(\omega_\nu)
  = 4 \nu_H T
  \sum_{\Omega_k}\sum_{n,m}\hat{q}_{mn}^{x}\Sigma_{nm}^{(e,Q)}
  \hat{q}_{nm}^{y}L_{n}(\Omega_k),
\label{DOS}
\\
M^z
= - \frac{\partial}{\partial H} \nu_H T
\sum_{\Omega_k}\sum_n \ln L_n^{-1}(\Omega_k).
\label{Mag}
\end{gather}%
Here
$L_n(\Omega_k)=
-\nu^{-1} \left[ \ln(T/T_{c})+\psi _{n}(|\Omega _{k}|)-\psi(1/2) \right]^{-1}$
is the fluctuation propagator,
$\psi_n(\Omega)$ is a short-hand notation for
$\psi[1/2+(\Omega +\alpha_n)/4\pi T]$,
with $\alpha_n=(4eDH/c)(n+1/2)$ being the Landau spectrum,
$\nu_H=eH/\pi c$,
and the matrix elements of the momentum operator in the Landau basis
are given by
$\hat{q}^{x,y}_{mn}
=\sqrt{eH/c} \, {\binom{i}{1}}
\bigl( \sqrt{m}\,\delta _{m,n+1}\mp \sqrt{n}\,\delta_{n,m+1}\bigr)$.

The three-Green-function blocks with two cooperons and electric ($-e$) or
heat ($i\left[ \varepsilon _{l}+\varepsilon _{l+\nu }\right]/2$)
vertices,
$\mathcal{B}_{nm}^{\left( e\right) }\left( \Omega _{k},\omega_\nu\right)$
and
$\mathcal{B}_{nm}^{\left( Q\right) }\left( \Omega _{k},\omega_\nu\right)$,
and the six-Green-function block with three cooperons and
electric and heat vertices,
$\Sigma_{nm}^{(e,Q)}(\Omega_{k},\omega_\nu)$,
shown in Fig.~\ref{figa1} should be calculated for an impure metal exactly
(the important ingredients can be found, e.g., in Refs.~\cite{LV,VGL}).
Performing summation over the fermionic Matsubara energies
we find the following explicit expressions ($\omega_\nu\geq0$):
\begin{multline}
  B_{nm}^{(e)}(\Omega_k,\omega_\nu)
  =
  e \nu D
  \left[
    \frac{\psi_m(\omega _{\nu}+|\Omega _{k}|) - \psi_n(|\Omega_{k}|)}
      {\omega_\nu+\Landau_m-\Landau_n}
  \right.
\\
  \left. {}
  + \frac{\psi_n(\omega _{\nu}+|\Omega _{k+\nu}|) - \psi_m(|\Omega_{k+\nu}|)}
      {\omega_\nu-\Landau_m+\Landau_n}
  \right] ,
\label{Be}
\end{multline}
\vspace{-7mm}
\begin{multline}
  B_{nm}^{(Q)}(\Omega_k,\omega_\nu)
  =
  \frac{-i\nu D}{2}
\\ {}
\times
\left[
  \frac{(\Omega_{k}-\Landau_m)\psi_m(|\Omega_k|+\omega_\nu)
    - (\Omega_{k+\nu}-\Landau_n)\psi_n(|\Omega_k|)}
    {\omega_\nu+\Landau_m-\Landau_n}
 \right.
\\
  \left. {}
  +
  \frac{(\Omega_{k+\nu}+\Landau_n)
    \psi_n(|\Omega_{k+\nu}|+\omega_\nu)
    -
    (\Omega_k+\Landau_m)\psi_m(|\Omega_{k+\nu}|)
    }
    {\omega_\nu+\Landau_n-\Landau_m}
\right] ,
\label{BQ}
\end{multline}
\begin{multline}
  \Sigma_{nm}^{\left(e,Q\right) }(\Omega_k,\omega_\nu)
  =
  - i e \nu {D}^2
  \left[
    \frac{\Omega_{k+\nu}-\alpha_n}{\omega_\nu+\alpha_m-\alpha_n}\psi_n'(|\Omega_k|)
  \right.
\\
  - \frac{\Omega_{k+\nu}+\alpha_n}{\omega_\nu-\alpha_m+\alpha_n}\psi_n'(|\Omega_{k+\nu}|+\omega_\nu)
\\
  -\frac{\Omega_k-\alpha_m}{(\omega_\nu+\alpha_m-\alpha_n)^2}\left(\psi_m(|\Omega_k|+\omega_\nu)-\psi_n(|\Omega_k|)\right)
\\
  \left.{}
  +\frac{\Omega_k+\alpha_m}{(\omega_\nu-\alpha_m+\alpha_n)^2}\left(\psi_n(|\Omega_{k+\nu}|+\omega_\nu)-\psi_m(|\Omega_{k+\nu}|)\right)\right]
    .
\label{Sigma}
\end{multline}

\begin{figure}[b]
\includegraphics[width=.95\columnwidth]{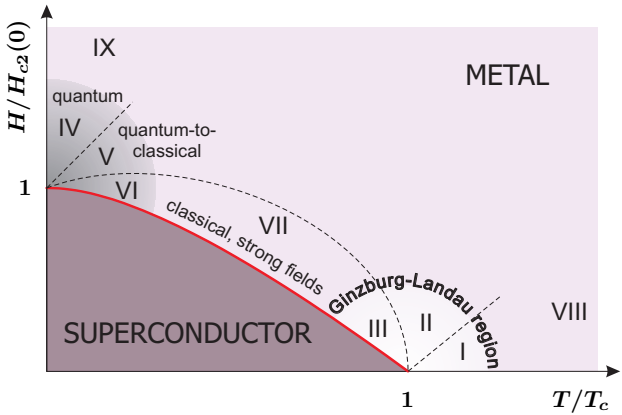}
\caption{(Color online) Different asymptotic regions for the fluctuation Nernst coefficient
on the $H-T$ phase diagram.}
\label{F:H-T-plane}
\end{figure}

The general calculation of Eqs.~(\ref{AL})--(\ref{Mag}) is
straightforward but very cumbersome.
However, one can identify nine
qualitatively different regions on the phase diagram (Fig.~\ref{F:H-T-plane}),
where the asymptotic behavior has a simple analytical form. The
corresponding nine asymptotes are presented below.
Before proceeding to the details, we point out
that the resulting expression for the Nernst coefficient
appears to be universal: Contrary to the behavior
of fluctuation conductivity \cite{LV},
the function $\beta^{xy}(T,H)$ depends only
on $T/T_c$ and $H/H_{c2}(0)$
and does not depend on the elastic scattering time $\tau$.
This is the magnetization contribution, $\beta_M^{xy}$,
which regularizes the otherwise divergent (and thus $\tau$-dependent)
terms in $\tilde\beta^{xy}$.

We start with discussing \emph{the classical regime close to the critical
temperature $T_{c}$}: The regions I, II, III in Fig.~\ref{F:H-T-plane} are
characterized by $\epsilon =\ln (T/T_{c})\ll 1$ and
$\tilde{h}=H/\widetilde{H}_{c2}(0)\ll 1$.
These are domains in which only classical AL
fluctuations are important and the DOS contribution is less singular.
The AL contribution to $\tilde{\beta}^{xy}$ reads
\be
  \tilde{\beta}^{xy}
  =
  2\beta_0 F(x)/x ,
\label{tilde-beta-123}
\ee
where $x=\epsilon /2\tilde{h}$, $\beta_0=k_Be/\pi\hbar=6.68$ nA/K
is the quantum of thermoelectric conductance,
and the function $F(x)$ is given by Eq.~(\ref{F(x)}).
One can see that the functional dependence of the corresponding
contribution to the Nernst coefficient on $x$
coincides with the qualitative estimate (\ref{nernstmagn}).
The magnetization contribution to the observable $\beta^{xy}$
[see Eq.~(\ref{betagen})] is given by
\be
  \beta_\text{M}^{xy}
  =
  \beta_0 \Big[\ln \frac{\Gamma (1/2+x)}{\sqrt{2\pi}}-x\,\psi (1/2+x)+x\Big] .
\label{beta-M-123}
\ee

In the limit of vanishingly small magnetic fields $\tilde{h}\ll \epsilon$
(region I), we find $\tilde\beta^{xy}=\beta_0(\tilde{h}/2\epsilon)$,
which is two times larger than the result
of Refs.~\cite{Uss1,Uss2,LV}.
The origin of the additional factor should be traced back to the
complicated analytic structure of the heat-current block (\ref{BQ}).
Thus the magnetization contribution
$\beta_\text{M}^{xy} = -\beta_0(\tilde{h}/6\epsilon)$
cancels only 1/3 of $\tilde\beta^{xy}$, and the final result
appears to be four times larger than the result of Refs.~\cite{Uss1,Uss2,LV}:
\be
  \beta_\text{I}^{xy}
  =
  \beta_0\frac{\tilde{h}}{3\epsilon }
  = \beta_0 \frac{\pi eDH}{12c(T-T_{c})},\qquad
\tilde{h}\ll \epsilon \ll 1.
\ee

In the limit $\epsilon \ll \tilde{h}$ (region II),
and close to the transition line,
at $\tilde{h}+\epsilon \ll \tilde{h}$ (region III),
Eqs.~(\ref{tilde-beta-123}) and (\ref{beta-M-123}) yield
\begin{gather}
  \beta_\text{II}^{xy}
  =
  \beta_0 \left[ 1- (\ln 2)/2 \right] ,
\qquad
  \epsilon \ll \tilde{h}\ll 1;
\\
  \beta_\text{III}^{xy}
  =
  \beta_0\frac{\tilde{h}}{\epsilon +\tilde{h}}
  =
  \beta_0\frac{H_{c2}\left( T\right) }{%
H-H_{c2}\left( T\right) },
\qquad
  \epsilon +\tilde{h}\ll \tilde h\ll 1.
\label{beta-above-Hc2(Tc)}
\end{gather}

Now we turn to the \emph{quantum regime close to the upper critical field}
$H_{c2}(0)=\pi cT_{c}/2\gamma eD$
(regions IV, V, VI in Fig.~\ref{F:H-T-plane}),
where $\gamma=1.78\dots$ is the Euler constant.
Here the role of magnetization term becomes crucial:
The $1/T$ divergence of $\beta_\text{M}^{xy}=cM^{z}/T$ exactly cancels the
divergent contribution originating from $\tilde\beta^{xy}$,
which is necessary to satisfy the
third law of thermodynamics. As a result, the total coefficient
$\beta_\text{IV}^{xy}$ remains finite in the limit or zero temperature.
The exact analytical expression at $t=T/T_{c}\ll 1$ and $\eta
=(H-H_{c2}(t))/H_{c2}(t)\ll 1$ is quite lengthy. We present below only the
asymptotic expressions in the regions IV, V, VI.

In the purely quantum limit of vanishing temperature and away from $H_{c2}(0)$
($t\ll \eta$, region IV), $\beta _{xy}$ is \emph{negative}:
\be
  \beta_\text{IV}^{xy}
  =
  - \frac{2\beta_0\gamma t}{9\eta}
  =
  - \frac{\beta_0 \pi c T}{9eD[H-H_{c2}(0)]},
\qquad
  t\ll \eta \ll 1.
\ee
This change of sign in thermoelectric response is similar to negative
magnetoresistance in the quantum fluctuation transport for the usual
electrical conductivity found in Ref.~{\cite{VGL} in the vicinity of $H_{c2}(0)$.
The sign change is due to
the DOS contribution being numerically larger than the positive AL term.

In the quantum-to-classical crossover region, where $H$ tends to $H_{c2}(t)$
but remains limited as $t^{2}/\ln (1/t)\ll \eta \ll t$ (region V), the
coefficient $\beta _{xy}$ becomes positive:
\be
  \beta_\text{V}^{xy}
  =
  \beta_0 \ln(t/\eta)
,\qquad t^{2}/\ln (1/t)\ll \eta \ll t\ll 1.
\label{beta-log-Hc2(0)}
\ee

Near $H_{c2}(t)$ ($\eta \ll t^{2}/\ln (1/t),$ region VI), we find:
\be
  \beta_\text{VI}^{xy}
  =
  8 \beta_0 \gamma^{2}t^{2}/3\eta ,
  \qquad
  \eta \ll t^{2}/\ln (1/t) \ll 1.
\label{beta-above-Hc2(0)}
\ee

We also address the full classical region \emph{just above the transition
line}, which covers a wide range of temperatures and magnetic fields
($\eta\ll 1$, region VII).
In this region, $\tilde{\beta}^{xy}$
is generally comparable to $\beta_\text{M}^{xy}=-\beta_0/\eta$,
and we obtain
\be
  \beta_\text{VII}^{xy}
  =
  \frac{\beta_0}{\eta }
  \left[
    1+\frac{h}{4\gamma t}\frac{\psi''(1/2+h/4\gamma t)}{\psi'(1/2+h/4\gamma t)}
  \right] ,
\qquad
  \eta\to0 ,
\label{beta-above-Hc2(T)}
\ee
with $h=H/H_{c2}(0)$. Close to $T_{c}$, Eq.~(\ref{beta-above-Hc2(T)})
matches Eq.~(\ref{beta-above-Hc2(Tc)}), while in the limit $T\rightarrow 0$
it matches Eq.~(\ref{beta-above-Hc2(0)}) provided that $\eta \ll t^{2}/\ln
(1/t)$. We note that in deriving Eq.~(\ref{beta-above-Hc2(T)}), the account
for Landau quantization of Cooper pair motion was crucial.

Finally, we address the \textquotedblleft non-singular\textquotedblright\
\emph{region far from the transition line} (regions VIII and IX in Fig.~\ref%
{F:H-T-plane}).
Similar to the fluctuation correction to conductivity, the Kubo
contribution $\tilde{\beta}^{xy}$ diverges as
$[\ln \ln (1/T_{c}\tau)-\ln \ln \max (h,t)]$, with $1/\tau$
playing role of the ultra-violet cutoff of the cooperon modes.
Remarkably, exactly the same divergence
of the opposite sign is contained in the magnetization
contribution $\beta_\text{M}^{xy}$. Therefore, the
total expression for $\beta^{xy}$ in this
region is convergent and decreases nearly as a power law when increasing $T$ and
$H$:
\begin{gather}
  \beta_\text{VIII}^{xy}
  =
  \beta_0
  \frac{eDH}{6\pi c T\ln (T/T_{c})},
  \qquad
  (1,h)\ll t ;
\label{beta-high-T}
\\
  \beta _{\text{\textrm{IX}}}^{xy}
  =
  \beta_0
  \frac{\pi c T}{12eDH\ln [H/H_{c2}(0)]},
  \qquad
  (1,t)\ll h.
\label{beta-high-H}
\end{gather}%
We see that even far from the transition the fluctuation Nernst coefficient
can be comparable or parametrically larger than the Fermi liquid terms. In
fact, it is conceivable that in some materials the Cooper channel
contribution to thermal transport at low temperatures can be dominant even
in the absence of any superconducting transition at all (e.g., if
superconductivity is \textquotedblleft hidden\textquotedblright\ by another
order).

\begin{figure}
\includegraphics[width=.95\columnwidth]{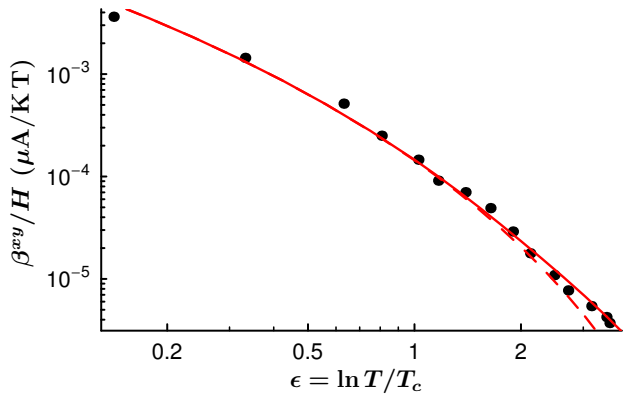}
\caption{(Color online) Comparison with experiment.
Circles: experimental data for $\lim_{H\to0}\beta^{xy}/H$
vs.\ $\epsilon=\ln T/T_c$ obtained for the 12.5-nm-thick
Nb$_{0.15}$Si$_{0.85}$ film~\cite{Aubin1}.
Dashed line: theoretical prediction for the strictly 2D geometry.
Solid line: theoretical prediction for the real sample \cite{Aubin1}
with the 2D-3D crossover taken into account.
}
\label{F:teor-and-exp}
\end{figure}

In Fig.~\ref{F:teor-and-exp},
we compare our theory with the experimental data \cite{Aubin1}
on the Nernst coefficient
for the Nb$_{0.15}$Si$_{0.85}$ film of thickness $d=12.5$ nm.
The dashed line shows our prediction for $\lim_{H\to0}\beta^{xy}/H$
in a wide range of temperatures up to $30\, T_c$
(including regions I and VIII).
We used the diffusion coefficient $D=0.087$ cm$^2$/s
which is 60\% of that reported in Ref.~\cite{Aubin1}
(with $k_Fl\sim1$, the precise determination of $D$
is questionable).
The data can be described by the 2D theory only in a close
vicinity of $T_c$, at $\epsilon\lesssim2$. For larger $\epsilon$,
the superconducting coherence length $\xi(T)$ becomes shorter
than $d$ and 3D nature of diffusion in the film should be taken into account.
This can be done by substituting $\Landau_n\to\Landau_n+D(\pi p/d)^2$
and performing an additional summation over $p=0,1,\dots$
in Eqs.~(\ref{AL})--(\ref{Mag}).
The resulting curve is shown in Fig.~\ref{F:teor-and-exp} by the solid line.

In summary, we have developed a complete microscopic theory of the
fluctuation Nernst effect in a two-dimensional superconductor. Our results
provide a natural explanation for a large Nernst signal observed in
superconducting films~\cite{Aubin1,Aubin2} and probably
should be relevant to cuprates,
where the energy scale of the chemical potential of preformed Cooper pairs
is set by the pseudogap temperature $T^*$.
Our theoretical predictions include a
near-power decay of the transverse thermoelectric response away
from the transition line,
which is expected to persist well into the metallic phase.

We are grateful to Herve Aubin, Mikhail Feigel'man and
Alexei Kavokin for useful discussions. M.N.S
acknowledges partial support from Dynasty Foundation and
hospitality of the University Paris-Sud.
V.G. acknowledges BU visitors program's hospitality. The work of M.N.S. and
M.A.S. was partially supported by RFBR Grant No.\ 07-02-00310.


\bibliography{Nernst}

\begin{thebibliography}{27}
\bibitem[{foo()}]{footnote}
\bibinfo{howpublished}{It is assumed that the impurities' potential is
  $\delta$-correlated in space so that
  $\tau(\epsilon)\nu(\epsilon)=\text{const}$.}
\end{thebibliography}

\end{document}